\documentclass[preprints,article,accept,moreauthors,pdftex]{mdpi}
\firstpage{1} 
\pubvolume{xx}
\issuenum{1}
\articlenumber{5}
\pubyear{2020}
\copyrightyear{2020}
\history{Received: date; Accepted: date; Published: date}

\Title{Towards personalized diagnosis of Glioblastoma in Fluid-attenuated inversion recovery (FLAIR) by topological interpretable machine learning.}

\Author{Matteo Rucco $^{1,\dagger,\ddagger}$\orcidA{}, Giovanna Viticchi $^{2,\ddagger}$ and Lorenzo Falsetti $^{3,\ddagger}$*}

\AuthorNames{Matteo Rucco, Giovanna Viticchi MD and Lorenzo Falsetti MD}

\address{%
$^{1}$ \quad United Technology Research Center, Via Praga 5, 38121 Trento (Tn), Italy.; matteo.rucco@utrc.utc.com\\
$^{2}$ \quad Neurological Clinic, Marche Polytechnic University, Ancona (An), Italy; Giovanna.Viticchi@ospedaliriuniti.marche.it\\
$^{3}$ \quad Internal and Sub-intensive Medicine Department, A.O.U. ``Ospedali Riuniti'', Ancona, Italy; Lorenzo.Falsetti@ospedaliriuniti.marche.it}

\corres{Correspondence:  matteo.rucco@utrc.utc.com; Tel.:  +39-0461-173-2317}

\firstnote{Current address: Raytheon Technology Research Center, Via Praga 5, 38121 Trento (Tn), Italy} 
\secondnote{These authors contributed equally to this work.}

\abstract{Glioblastoma multiforme (GBM) is a fast-growing and highly invasive brain tumour, it tends to occur in adults between the ages of 45 and 70 and it accounts for 52 percent of all primary brain tumours. Usually, GBMs are detected by magnetic resonance images (MRI). Among MRI, Fluid-attenuated inversion recovery (FLAIR) sequence produces high quality digital tumour representation. Fast detection and segmentation techniques are needed for overcoming subjective medical doctors (MDs) judgment. In the present investigation, we intend to demonstrate by means of numerical experiments that topological features combined with textural features can be enrolled for GBM analysis and morphological characterization on FLAIR. To this extent, we have performed three numerical experiments. In the first experiment, Topological Data Analysis (TDA) of a simplified 2D tumour growth mathematical model had allowed to understand the bio-chemical conditions that facilitate tumour growth: the higher the concentration of chemical nutrients the more virulent the process. In the second experiment topological data analysis was used for evaluating GBM temporal progression on FLAIR recorded within 90 days following treatment (e.g.,  chemo-radiation therapy - CRT) completion and at progression. The experiment had confirmed that persistent entropy is a viable statistics for monitoring GBM evolution during the follow-up period. In the third experiment we had developed a novel methodology based on topological and textural features and automatic interpretable machine learning for automatic GBM classification on FLAIR.  The algorithm reached a classification accuracy up to the 97\%.}


\begin{document}

\section{Introduction}
\label{sec:introduction}
Gliomas are the most common primary brain tumours, originating from glial cells. We can differentiate between benign tumours, with in some cases a lifelong expectancy, and malignant forms. In this second group, glioblastoma multiforme (GBM) is the most frequent malignant cancer with the worst prognosis, with less than 5\% of affected patients with a 5-year survival rate and with the highest relapse rate~\cite{batash2017glioblastoma}. GBM are characterized by a diffuse and infiltrative growth pattern, as invasive glioma cells often migrate along myelinated white matter (WM) fiber tracts. This is a major cause of their appalling prognosis: tumour cells invade, displace, and possibly destroy WM. For these reasons, surgical, radiotherapic and chemotherapic approaches, also early performed, rarely resulted effective for a significant number of months~\cite{weller2014european}. Moreover, large surgical resections could cause unacceptable effects on motor or speech functions. In largely infiltrative GBM the real objective of surgical approach is to leave less possible cancer cells in situ to give more chances to adjuvant therapy. The non-invasive detection of microscopic infiltration, as well as the identification of aggressive tumour components within spatially heterogeneous lesions, are of outstanding importance for surgical and radiation therapy planning or to assess response to chemotherapy. Moreover, the critical importance of an accurate detection of local infiltration is underlined by the results of intraoperative MR: this technique seems to be able to contribute significantly to optimal resection~\cite{kuhnt2011correlation} and to improve post-operative outcomes~\cite{schijns2011oe}.Magnetic resonance (MR) imaging plays an important role in the detection and evaluation of brain tumours. The evaluation of post-surgical residual cancer by MR should be performed in all patients after 48-72 hours after surgery, because this factor has a relevant prognostic value both for survival and for the subsequent therapeutic response~\cite{majos2016early}. For more than 30 years~\cite{damadian1971tumour} conventional MR imaging, typically T1w images before and after paramagnetic contrast administration, T2w images and FLAIR images have been largely used to evaluate brain neoplasms.  MR allows to localize the lesions, helps to distinguish tumours from other pathological processes, and depicts basic signs of response to therapy, such as changes in size and degree of contrast enhancement.  Manual tumour detection and segmentation on MR images is time-consuming and can have a great inter-observer variability; automatic segmentation is more reproducible and efficient when robustness is particularly desirable, such as in monitoring disease progression or in the longitudinal evaluation of emerging therapies~\cite{castellano2016evaluation}. This is a very relevant element because often in clinical practice the decisions regarding therapy continuation or discontinuation are taken on the basis of disease recovery ~\cite{weller2014european}. Radiomics is a set of techniques for extracting a vectorial representation that provides a quantitative description of tumour radiographic data. In general, radiomics extracts quantitative description of signals and images intensity, texture description and geometric characteristics.
Texture analysis, an image-analysis technique that quantifies grey-level patterns by describing the statistical relationships between the intensity of pixels, has demonstrated considerable potential for cerebral lesion characterization and segmentation~\cite{castellano2004texture,remondino2006image,schad1993ix}. For automatic glioma detection and segmentation in MRI, several algorithms have been already proposed. The most recent works can be grouped into superpixel based segmentation and deep learning based segmentation. In~\cite{zhao2018automated}, the authors presented an interesting bottom-up approach that aims to combine graphical probabilistic model (i. e, Conditional Random Fields - CRF -) for capturing the spatial interactions among image superpixel regions and their measurements. A number of features including statistics features, the combined features from the local binary pattern as well as grey level run length, curve features, and fractal features were extracted from each superpixel. The features were used for teaching machine learning models to discriminate between healthy and pathological brain tissues. In~\cite{wu2019automatic}, the authors used the concept of superpixels based segmentation for instrumenting a machine learning algorithm (i.e., support vector machine - SVM) that uses both textural and statistical features. The authors reported excellent average Dice coefficient, Hausdorff distance, sensitivity, and specificity scores when applying their method on T2w sequences from the Multimodal Brain tumour Image Segmentation Benchmark 2017 (BraTS2017) dataset. In~\cite{shivhare2020brain}, the authors exploited a new method for brain tumour detection by combining features computed from Fluid-Attenuated Inversion Recovery (flair) MRI in association with the graph-based manifold ranking algorithm. The algorithm counts three mains steps: in the first phase, superpixel method is used to convert the homogeneous pixels in the form of superpixels. Rank of each superpixel or node is computed based on the affinity against certain selected nodes as the background prior in the second phase. The relevance of each node with the background prior is then computed and represented in the form of tumour map. In~\cite{soltaninejad2017automated} the authors have reported on an approach that computes for each superpixel a number of novel image features including intensity-based, Gabor textons, fractal analysis and curvatures within the entire brain area in FLAIR MRI to ensure a robust classification. The authors compared Extremely randomized trees (ERT) classifier with support vector machine (SVM) to classify each superpixel into tumour and non-tumour. 
Deep learning based solutions are becoming the new tools for brain segmentation. In~\cite{korfiatis2016automated},the authors used autoencoders for instrumenting an automatic segmentation of increased signal regions in fluid-attenuated inversion recovery magnetic resonance imaging images. In~\cite{lorenzo2019segmenting}, the authors have provided a solution for dealing to the limited amount of available data from ill brains. They have trained a one-class classifier algorithm based on deep learning for segmenting brain tumours from fluid attenuation inversion recovery MRI. The technique exploits fully convolutional neural networks, and it is equipped with a battery of augmentation techniques that make the algorithm robust against low data quality, and heterogeneity of small training sets. Beside segmentation, deep-learning based solutions have been exploited for skull-stripping, and tissues identification (white matter, grey matter, etc. . )~\cite{kleesiek2016deep, galdames2012accurate}, 
At the best of our knowledge, latest radiomics approaches are reported in~\cite{chaddad2019radiomics}.
Another approach for the analysis of cancer development is based on dynamical system theory~\cite{bellomo2000modelling}. The models differ each other for different degrees of abstraction of the environment in which the tumour growth. Usually, the environment is modeled by the parameters in the equations, and they describe both the chemical nutrients and the spatial constraints. We notice that most of the models approximate tumour cells shape with spheres. Discrete models such as agent-based models reproduce the evolution of single spheroids. Partial Differential equation such as reaction-diffusion equation are used for predicting the temporal and spatial evolution of tumour cells. They can account for numerous biological phenomenon, but they are very sensitive to model parameters, making them hardly fittable with quantitative biological data in simple experimental set-up~\cite{michel2018mathematical}.
Recent works have combined radiomics with dynamical system for providing personalized model. A complete and exhaustive mathematical modelling of tumour growth is quite impossible since it involves too many subsystems which have an effect in the growth of a tumour. In~\cite{unsal2019personalized}, the authors have defined a novel approach for identifying the fundamental components involved in cancer development. They have proposed a novel probability functions to obtain a personalized model and estimate the individual importance of each subsystem by means of simulated annealing for parameter optimization. The authors have validated the prediction of tumour growth with in-vitro tumour growth data.

\subsubsection*{Topological data analysis for radiomics and tumour growth analysis}
Topological space is a powerful mathematical concept for describing the connectivity of a space. Informally, \textit{a topological space} is a set of points, each of which equipped with the notion of \textit{neighboring}~\cite{hatcher2002algebraic, munkres1984elements}. In the last decade a new suite of tools, based on algebraic topology, for data exploration and modelling haven been invented~\cite{carlsson,zomorodian2007topological,edelsbrunner2008persistent}. The data science community refers to these tools as \textit{Topological Data Analysis} (TDA). TDA has been used in different domains: biology, manufacturing, medicine and others~\cite{rucco2016survey}.  Topological entropy, namely Persistent Entropy, is equipped with suitable mathematical properties, that permits to describe complex systems~\cite{atienza2019persistent} and it has been applied in different experiments, e.g. the analysis of biological images~\cite{jimenez2017topological} and the analysis of medical signals~\cite{piangerelli2018topological}.
At the best of our knowledge, the extraction of topological features for radiomics from topological data analysis is still at its infancy.
In~\cite{oyama2019hepatic}, the authors have compared the accuracy of  machine learning models for the classification of hepatic tumours. From T1-weighted magnetic resonance (MR) images, the authors have computed both texture analysis and topological data analysis using persistent homology. The textural features or the topological features were used as input for machine learning models, the best accuracy (92\%) for hepatocellular carcinomas was obtained with textural features, while TDA based machine learning model obtained the 85\% of accuracy for metastatic tumours.
Recent papers have demonstrated that topological data analysis algorithms are suitable for the analysis of dynamical systems~\cite{gholizadeh2018short}. A common approach is to embed the time series of a dynamical system into a point cloud data and to study its shape by means of persistent homology.  In~\cite{myers2019persistent}, the authors have demonstrated that Persistent Entropy, a topological feature, is able to distinguish between periodical and chaotic systems. An alternative approach is represented by the analysis of the phase space of the system by means of TDA~\cite{camara2017topological}. 
In particular, we are interested to detect and to study topological loops (1D, 2D, etc...) that would surround brain regions of interest. We have introduced a novel entropy, the so-called \textit{generator entropy}, that is a function of the number of 0-simplices (vertices) in the loops. Our intuition is that a pathological brain is characterized by the presence of a mass with a different grey gradient that would correspond to a topological loop. If the mass structure is heterogeneous or if the brain is affected by multiple tumours the number of loops will be higher. An advantage of using persistent homology as radiomics features on FLAIR is that one can detect the differences between the shapes associated to pathological and healthy brain tissues. 

\subsubsection*{Paper outline}
In the present investigation, we intend to demonstrate by means of numerical experiments that topological features combined with GLCM features can be enrolled for GBM analysis and morphological characterization on FLAIR. To this extent, we have performed the following numerical experiments:
\begin{enumerate}
    \item Analysis 1 - Topological Data Analysis of a simplified 2D tumour Growth Mathematical Model: identification how tumour growth over time is affected by the initial amount of available chemical nutrient
    \item Analysis 2 - Topological analysis of Glioblastoma temporal progression on FLAIR: evaluation of GBM temporal evolution after treatment 
    \item Analysis 3 - Automatic GBM  classification on FLAIR: the aims of this experiment is to evaluate the accuracy for classification GBM by characterization of 2D patches extracted from FLAIR by combining textural and topological data analysis with machine learning. 
\end{enumerate}

In the first experiment, we have integrated a discrete time mathematical model to study the evolution of three different type of tumour cells: proliferative, quiescent and necrosis. At each time step, we associated a point cloud data (PCD) to each cell set. The spatial evolution of the PCD is analysed by topological statistics, i.e. persistent entropy. The plot of persistent entropy over time reveals interesting system's evolution. The results encouraged us to analyse if topological features can detect also GBM temporal evolution. To this extent, we have analysed a publicly available dataset that contains 2 sequences of FLAIR for 20 patients. The sequences were recorded within 90 days following chemo-radiation therapy (CRT) completion and at progression. The complete description of the dataset is in Sec.:~\ref{sec:discussion}. For the sake of completeness, we decided to use FLAIR since they reveal a wide range of lesions, including cortical, periventricular, and meningeal diseases that were difficult to see on conventional images. 
In the third experiment we have used the same dataset but for evaluating the possibility of discriminating healthy and pathological tissue on FLAIR, by the use of Statistical Texture Analysis and Topological Data Analysis. The discriminating power of statistical texture and of topological features was then exploited for the development of a supervised tumour detection methodology by means of automatic machine learning algorithms. We have adopted novel computational tools for debugging and understanding Machine Learning decision.   

The paper is organized as follows: in Section~\ref{sec:background} we introduce the relevant background, namely the fundamental concepts of \textit{textural features} and \textit{topological features}. Section~\ref{sec:methodology} introduces the  methodology for GBM characterization. In Section~\ref{sec:discussion} we report on the implementation and application of the methodologies on slices extracted from public domain FLAIR. Final thoughts about the results and next steps are discussed in the last Section~\ref{sec:conclusion}.

\section{Materials}
\label{sec:materials}
\subsection*{Analysis 1: Topological Data Analysis of a simplified 2D tumour Growth Mathematical Model}
\label{sec:an1}
In order to investigate the topological properties of tumour growth dynamics, we have used a well documented tumour growth model, the so-called Sherratt-Chaplain model~\cite{sherratt2001new}. The model is also valid for GBM like tumour~\cite{bottger2012investigation}. The model describes the temporal evolution of cell densities of profilerating, quiescent and necrotic cells~\cite{stein2007mathematical}. Cells movements (diffusion) is modeled by a competitive approach. The model has been updated and compared with in-vitro analysis for a fine-tuning of its parameters~\cite{ ang2009analysis}. The motion equation for the profilerating $p$, quiescent $q$ and necrotic cells $n$ is:
\begin{equation}
    \frac{dp}{dt}=\frac{\partial}{\partial x}(\frac{p}{p+q}\frac{\partial(p+q)}{\partial x} )+g(c)p(1-p-q-n)-f(c)
\end{equation}
\begin{equation}
    \frac{dq}{dt}=\frac{\partial}{\partial x}(\frac{q}{p+q}\frac{\partial(p+q)}{\partial x} )+f(c)p-h(c)q
\end{equation}
\begin{equation}
    \frac{\partial n}{\partial t}=h(c)q
\end{equation}
Where $c$ is the concentration of some nutrients that can be represented by:
\begin{equation}
    c=\frac{c_0\gamma}{\gamma + p}(1-\alpha(p+q+n))
\end{equation}
The functions $f,g$ and $h$ are defined:
\begin{equation}
    f(c)=\frac{1-tanh(4c-2)}{2}
\end{equation}
\begin{equation}
    g(c)=\beta e^{\beta c}
\end{equation}
\begin{equation}
    h(c)=\frac{f(c)}{2}
\end{equation}
The model can be solved with numerical methods by translating them into finite differences~\cite{ang2009analysis}. Under the assumption of radial symmetry the displacement of the cells can be seen 2-dimensional.  We remark that we do not intend to deep dive in mathematical modelling of GBM growth, but we want to investigate topological properties of GBM like tumour.

\subsection*{Analysis 2 \& 3: Dataset}
In order to make our results comparable, the analysis 2 and 3 were executed by using a public freely available dataset accessible via \textit{The Cancer Imaging Archive (TCIA)}~\cite{clark2013cancer}.  In details, the dataset includes DICOM files of 20 subjects from different sites with primary newly diagnosed glioblastoma who were treated with surgery and standard concomitant chemo-radiation therapy (CRT) followed by adjuvant chemotherapy. The sequences are T1-weighted (pre and post-contrast agent), FLAIR, T2-weighted, ADC, normalized cerebral blood flow, normalized relative cerebral blood volume, standardized relative cerebral blood volume, and tumour masks (i.e., ROIs)~\cite{schmaindakmprahm2018,ellingson2009functional}. Fig.~\ref{img:Slices} shows two among the 2D slices contained in the dataset and extracted from FLAIR. Each patient is described by two MRI exams: within 90 days following CRT completion and at progression. At the best of our knowledge, in this paper the dataset is used for the first time for ML experiments~\cite{schmainda2018multi}.  
The following preprocessing steps were performed before running the analysis:

\begin{figure}[!ht]
\centering
\includegraphics[width=8cm]{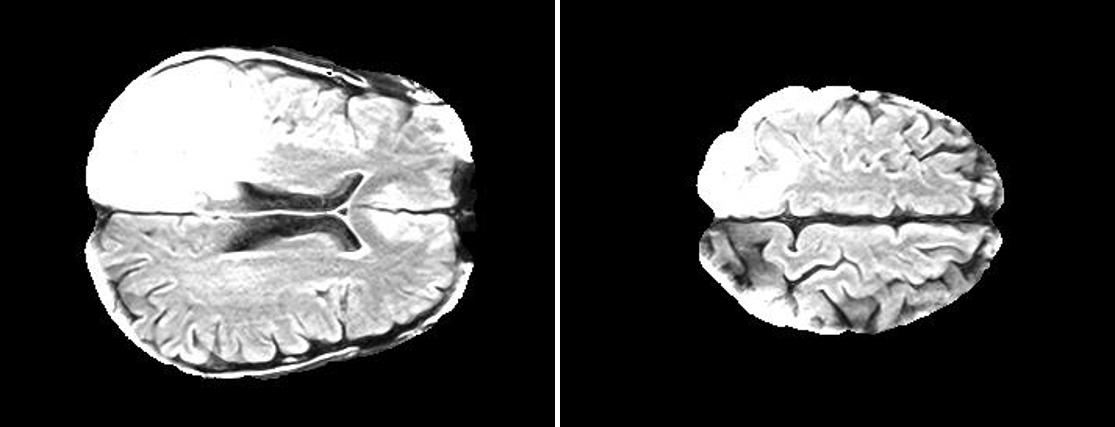}
\caption{Example of slices contained in the dataset.}
\label{img:Slices}
\end{figure}  

\subsection*{FLAIR preprocessing}
The following preprocessing steps attempt to enhance and correct FLAIR for posterior analysis. We propose the following scheme for the dataset under analysis: 
\begin{itemize}
    \item DICOMs were transformed into NII files for enabling preprocessing steps\footnote{\url{https://pypi.org/project/dicom2nifti/}}. 
    \item FLAIRs were optimized by removing fat tissues and by performing skull stripping. The preprocessing steps, namely the removal of fat tissues and skull stripping, were performed by using the deep-learning algorithm described in~\cite{lipkova2019personalized}\footnote{\url{https://github. com/JanaLipkova/s3}}.
\end{itemize}

\section{Background}
\label{sec:background}

\subsubsection*{Topological Data Analysis: persistent homology}
Homology is an algebraic machinery used for describing a topological space $\mathfrak{C}$. 
Informally, for a fixed natural number $k$, the $k-$\textit{Betti number} $\beta _k$ counts  the number of $k-$dimensional holes characterizing $\mathfrak{C}$: $\beta_0$ is the number of connected components, $\beta_1$ counts the number of
holes in 2D or tunnels in 3D\footnote{Here $n$D refers to the $n-$dimensional space $\mathbb{R}^n$. }, $\beta_2$ can be thought as the number of voids in geometric solids, and so on. 

Persistent homology is a method for computing the $k-$dimensional holes at different spatial resolutions. Persistent holes are more likely to represent true features of the underlying space, rather than artifacts of sampling (noise), or due to particular choices of parameters.  For a more formal description we refer the reader to~\cite{edelsbrunner2010computational}. In order to compute persistent homology, we need a distance function on the underlying space. This can be obtained constructing \textit{a filtration} on a simplicial complex, which is a nested sequence of increasing subcomplexes. More formally, a filtered simplicial complex $K$ is a collection of subcomplexes $\{K(t): t \in \mathbb{R}\}$ of $K$ such that $K(t) \subset K(s)$ for $t< s$ and there exists $t_{max}\in \mathbb{R}$ such that $K_{t_{max}}=K$. The filtration time (or filter value) of a simplex $\sigma \in K$ is the smallest $t$ such that $\sigma \in K(t)$.  A $k-$dimensional Betti interval, with endpoints $[t_{start}, t_{end}),$ corresponds to a $k-$dimensional hole that appears at filtration time $t_{start}$ and remains until time $t_{end}$. We refer to the holes that are still present at $t= t_{max}$ as \textit{persistent topological features}, otherwise
they are considered \textit{topological noise}~\cite{adams2011javaplex}. Figure~\ref{img:Persistent} depicts the computation of persistent homology on a FLAIR slice. The methodology for transforming the image into a weighted graph is described in the text. 
The set of intervals representing birth and death times of homology classes is called the {\it persistence barcode} associated to the corresponding filtration. 
Instead of bars, we sometimes draw points in the plane such that a point $(x,y)\in \mathbb{R}^2$ (with $x< y$) corresponds to a bar $[x, y)$ in the barcode. This set of points is called {\it persistence diagram}. There are several algorithms for computing persistent homology and their analysis and for a complete overview of the available tools we refer to~\cite{otter2017roadmap}.

\subsection*{Topological Features for Radiomics}
In order to use persistent barcodes in any computer applications (e.g. machine learning), it is needed to derive a set of numerical descriptors which encapsulate the information contained in the barcode. In the following we recall how to compute the statistics we use in this research.
Two or more topological objects, e.g. simplicial complexes, are homologically equivalent if they have the same sequence of Betti numbers. In other words by using persistent homology, they are characterized by the same number of persistent homological holes at each dimensioms. To facilitate the comparison of two or more simplicial complexes, one can calculate and compare the Euler Characteristics for each simplicial complex:

\begin{Definition}[Euler Characteristic]\ \\
$$
\chi = \sum_{i=0} ^{i=n} (-1)^{i} \beta_{i}
$$
Where $\beta_i$ is the Betti numbers at the \textit{i}-th homological group (e.g., $\beta_0$ is the Betti number at $H0$ for counting the number of connected components, $\beta_1$ for counting the number of 2D holes, $\beta_2$ for counting the number of 3D empty volumes etc\dots). 
\end{Definition}
Since Euler Characteristics is computed on the persistent homological holes and it discards completely the noisy topological features (i.e., not persistent) we suggest to complement it compute also the so-called Persistent Entropy. 
Persistent Entropy is a Shannon like entropy computed over the persistent barcodes and it is calculated by using both noisy and persistent topological features.. It was defined initially in ~\cite{chintakunta2015entropy} and further studies of its mathematical properties were published in ~\cite{rucco2015characterisation,rucco2017new}. We recall its definition.

\begin{Definition}[Persistent Entropy] \ \\
Given a persistence barcode $B = \{a_i=[x_i , y_i) : i\in I\}$, the \textit{Persistent Entropy} (PE) $H$ of the filtered simplicial complex is defined as follows:
$$
H=-\sum_{i \in I} p_i log_{10}(p_i)
$$
where $p_i=\frac{\ell_i}{L}$, $\ell_i=y_i - x_i$, and $L=\sum_{i\in I}\ell_i$. 
\end{Definition}

In the case of an interval with no death time, $[x_i,+\infty)$, we truncate infinite intervals and replace $[x_i,+\infty)$ by $[x_i~,~m)$ in the persistence barcode, where $m = t_\mathrm{max} + 1$. 

Note that the maximum PE corresponds to the situation in which all the intervals in the barcode are of equal length. In that case, $H=\log n$ if $n$ is the number of elements of $I$. Conversely, the value of the PE decreases as more intervals of different length are present. 
A topological $n$-dimensional hole is generated by the so-called \textit{homological generator}~\cite{ghrist2008barcodes,obayashi2018volume}. For example, a 1D hole is generated by a set of 0-simplices (vertices) linked by 1-simplices  (edges). Given our interest to $n$D topological holes, we propose a new seminal statistics that is computed on the number of 0D simplices (vertices) in each topological features. 
\begin{Definition}[Generator Entropy] \ \\
Given a filtered simplicial complex $\{K(t) : t\in F\}$, and the collection of corresponding persistence barcode  $B = \{a_i=[x_i , y_i) : i\in I\}$ for the Homological Groups $H_j$ with $j\geq1$.
$$
  GH = - \Sigma_{i = 1} ^ {i = N}(p_i log_{10}(p_i))~\newline
$$
where N is the total number of homological lines in the barcode, $n_i$ is the number of 0-simplices in the \textit{i}-th line and $L = \Sigma_{i = 1} ^{i = N} n_i$ and $p_i = \frac{n_i}{L}$.  For the sake of preciseness, in a topological loop there exist a 0-simplices that appears twice. In this setting we count each 0-simplices only once. 
\end{Definition}
We envision that this new topological statistics could be used as measure of number of different objects surrounded by topological loops and their diameter.

\subsection*{Textural features : grey-Level Co-occurrence matrix}
In this study we used one selected feature set (grey-Level Co-occurrence matrix, GLCM) for the texture analysis in accordance with some previous reports of radiomics for GBM detection~\cite{chaddad2015radiomics}. GLCM is a statistical method on examining image texture by taking into account the spatial relationship of pixels. The GLCM counts how often pairs of pixel with specific values and in a specified spatial relationship occur in an image. From GLCM we have extracted the following textural information~\cite{haralick1973textural}:
\begin{enumerate}
      \item Contrast - it measures the local variability in the grey level. 
      \item Correlation - it measures the joint probability that a given pair of pixels is found. 
      \item Homogeneity - it measures the distance between the GCLM diagonal and element distribution. 
      \item Energy
\end{enumerate}

In this paper we have used the open-source software Ripser~\cite{ctralie2018ripser}\footnote{\url{https://ripser.scikit-tda.org/}} for TDA. Ripser allows to compute persistent homology of both point cloud data and 2D images. For the computation of persistent homology we have used a slightlhy modified version of the lower star filtration \footnote{\url{ https://ripser.scikit-tda.org/notebooks/Lower Star Image Filtrations.html }}. In our version, the algorithm computes the homology up to the first homological group and it returns for each topological loop a set of representative generators. While for GLCM computation we have used Scikit-Image~\cite{van2014scikit}.

\section{Methods}
\label{sec:methodology}
\subsection{Analysis 1: Topological Data Analysis of a simplified 2D tumour Growth Mathematical Model}
We have integrated the model introduced in Sec.\ref{sec:an1} by using the code provided in the paper~\cite{ang2009analysis} but rewritten in Python. The dimensionless parameters $\alpha$, $\gamma$ and $\beta$ are extracted from in-vitro experiments. We remark that $\alpha$ and $\gamma$ represent concurr to the definition of the initial concentration of chemical nutrient. We have analysed how the system is affected by different values of $\alpha$ by keeping $\gamma = 10.0 $ and $\beta=0.5$. For each value of $\alpha$, and throughout the integration, the 2D coordinates of each cell were extracted. The coordinates were used as input for TDA. Fig.:\ref{img:distribution} depicts both the spatial distribution of the cells and the corresponding topological features computed throughout the system's integration.
\begin{figure}[!ht]
\centering
\includegraphics[width=14cm]{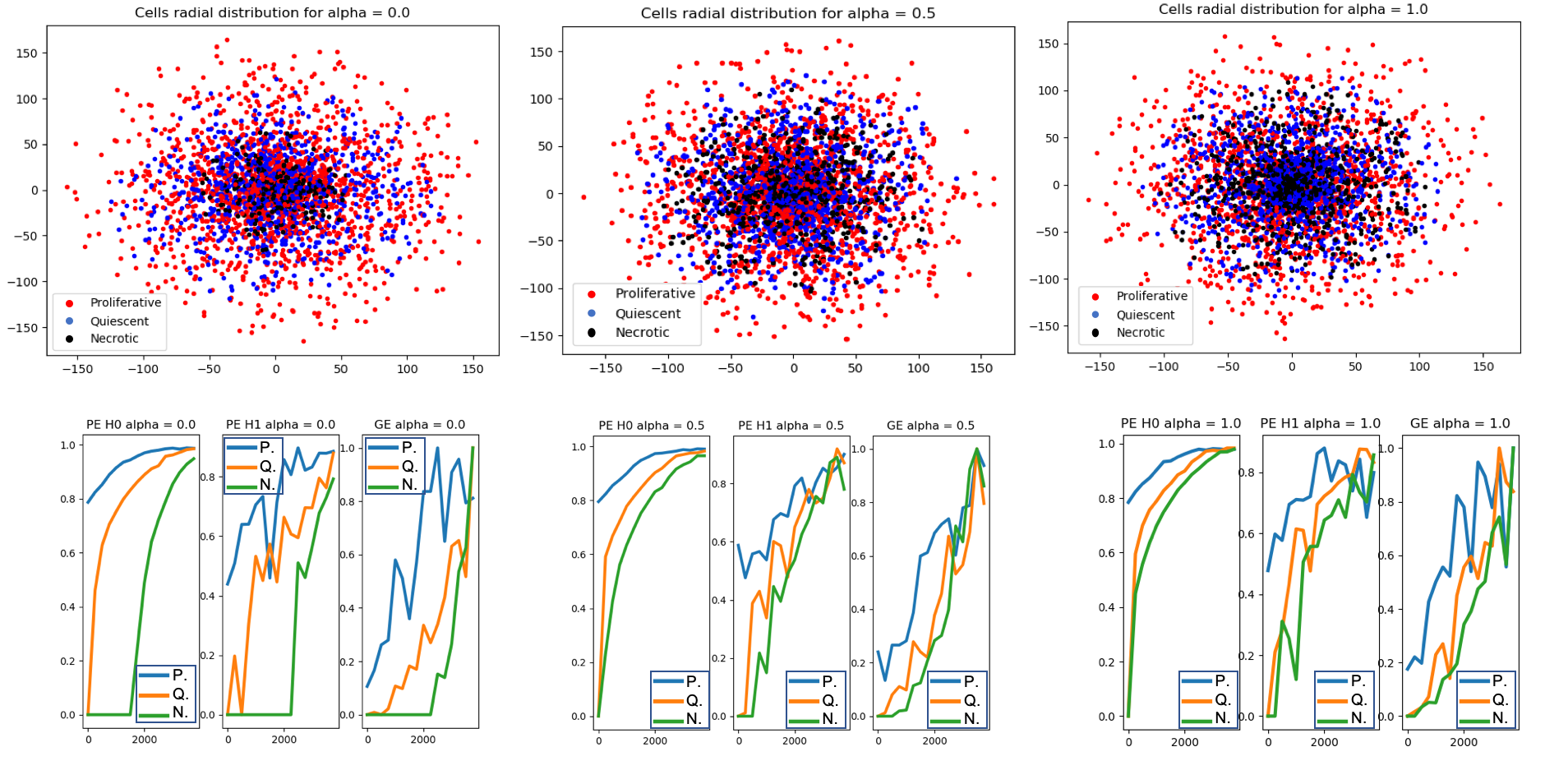}
\caption{Top: radial cell displacement for different $\alpha=[0.0,0.5,1.0]$. In red: proliferative, blue: quiescent, black: necrotic. Bottom: Corresponding temporal evolution of Persistent Entropy for H0 and for H1 and Generator Entropy for the same $\alpha$: for proliferative cells (blue), quiescent cells (orange) and necrotic cells (green) }
\label{img:distribution}
\end{figure}  

\subsection{Analysis 2: Topological Data Analysis of tumour Progression}
\label{sec:an2} 
\begin{figure}[!ht]
\centering
\includegraphics[width=12cm]{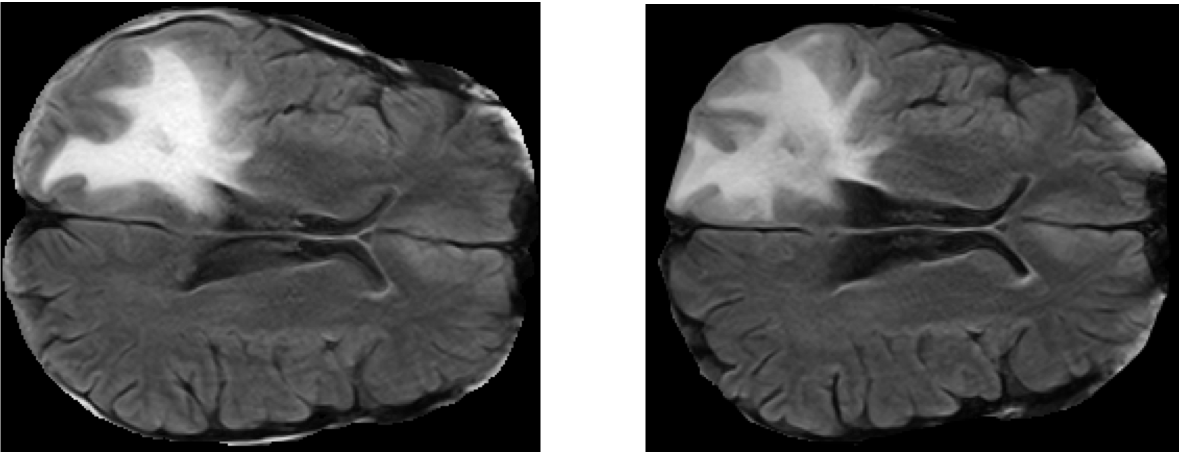}
\caption{Example of two slices from FLAIR for the same patient within 90 days following chemo-radiation therapy (CRT) completion and at progression}
\label{img:pre_post}
\end{figure}

The dataset contains for each patient two MRI exams: the exams were recorded within 90 days following CRT completion and at progression, for now on ''pre" and ''post". For each patient the topological features (i.e. Euler characteristics, persistent entropy at H0 and H1 and generator entropy) were computed on the 2D slices of FLAIR. Welch’s t-test is used for comparing the mean of the topological descriptor computed over the ''pre" and ''post" set of images. Fig.:\ref{img:pre_post} depicts of two FLAIR slices extracted for the same patients within 90 days following chemo-radiation therapy (CRT) completion and at progression.

\subsection{Analysis 3: GBM classification}
\label{sec:an3}
\begin{figure}[!ht]
\centering
\includegraphics[width=12cm]{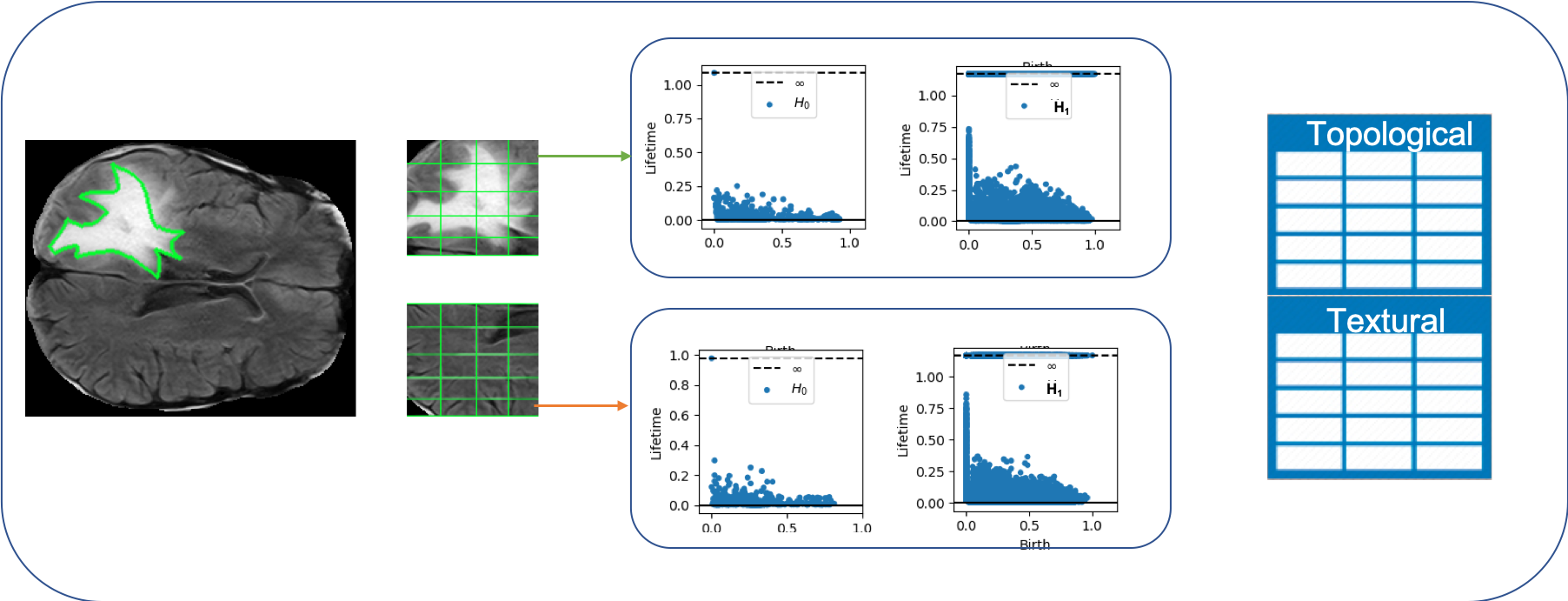}
\caption{Pictorial representation of the methodology for GBM classification. From left to right: 2D FLAIR slices with ROI surrounding GBM lesion. Lateral and controlateral ROIs divided in rectangular patches shaping a grid on both of them. Example of Persistent Barcodes for H0 (connected component) and H1 (2D topological loops). Table containing both topological and textural features used for subsequent analysis.)}
\label{img:process}
\end{figure}

In this work we have implemented and tested the following GBM classification methodology using texture and topological data analiysis. The input for the methodology are the FLAIR and the corresponding ROI files. For each 2D slice we have executed the following process that is also depicted in Fig.:\ref{img:process}:
\begin{enumerate}
\item 2D GLCM features and topological features were calculated with a sliding patch approach in the segmented ROI. The size of the patch is $30\times30$.
\item In order to identify discriminating features, the same feature calculations were also performed in the contralateral (healthy) ROIs. 
\item Each sliding patch was labeled as healthy or pathologic according to the class of its ROI. Each patch was also stored for future analysis.
\item Features selection.
\item The dataset is randomly divided in \textit{training} and \textit{testing} subsets that contain the 70\% and 30\% of samples respectively.
\item The training set is used for training a machine learning classifier. For the sake of completeness, during training we have adopted a k-fold cross validation standard procedure~\cite{wong2015performance}\footnote{\url{https://machinelearningmastery.com/difference-test-validation-datasets/}} with $k=5$. 
\item Splitting, training and testing procedure were executed multiple times by using different set of features: only topological features or only GLCM features or topological plus GLCM features.  
\item Classifier behavior is debugged by tools from information theory. This allows to understand feature relevance and to understand what are the numerical input characteristics related to the classification verdict. Specifically, we have used Skater and Lime algorithms~\cite{wei2015variable}\footnote{\url{https://www.oreilly.com/ideas/interpreting-predictive-models-with-skater-unboxing-model-opacity}}.
\end{enumerate}

\subsubsection{Feature selection}
\begin{figure}[!ht]
\centering
\includegraphics[width=8cm]{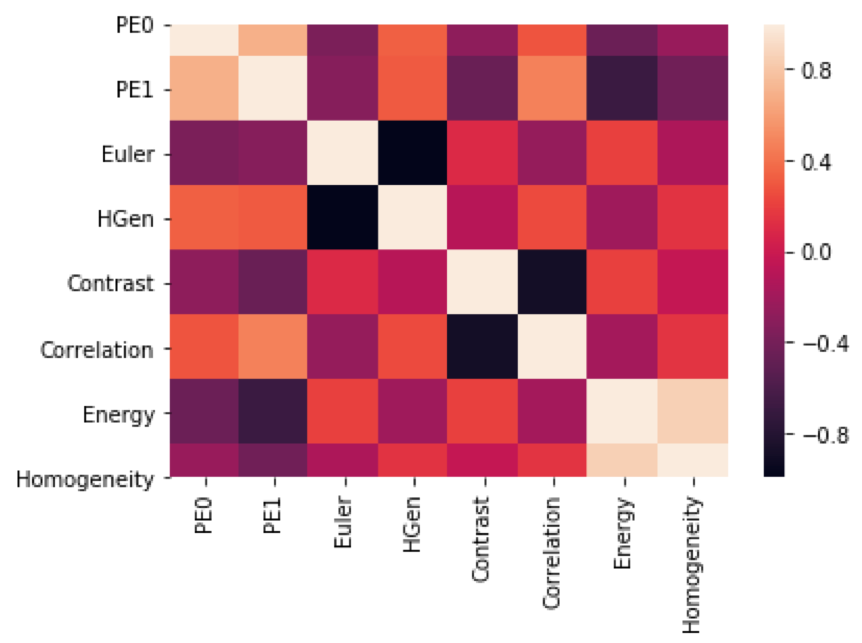}
\caption{Heat map depicting the correlations among variables.}
\label{img:correlation}
\end{figure}

\begin{figure}[!ht]
\centering
\includegraphics[width=14cm]{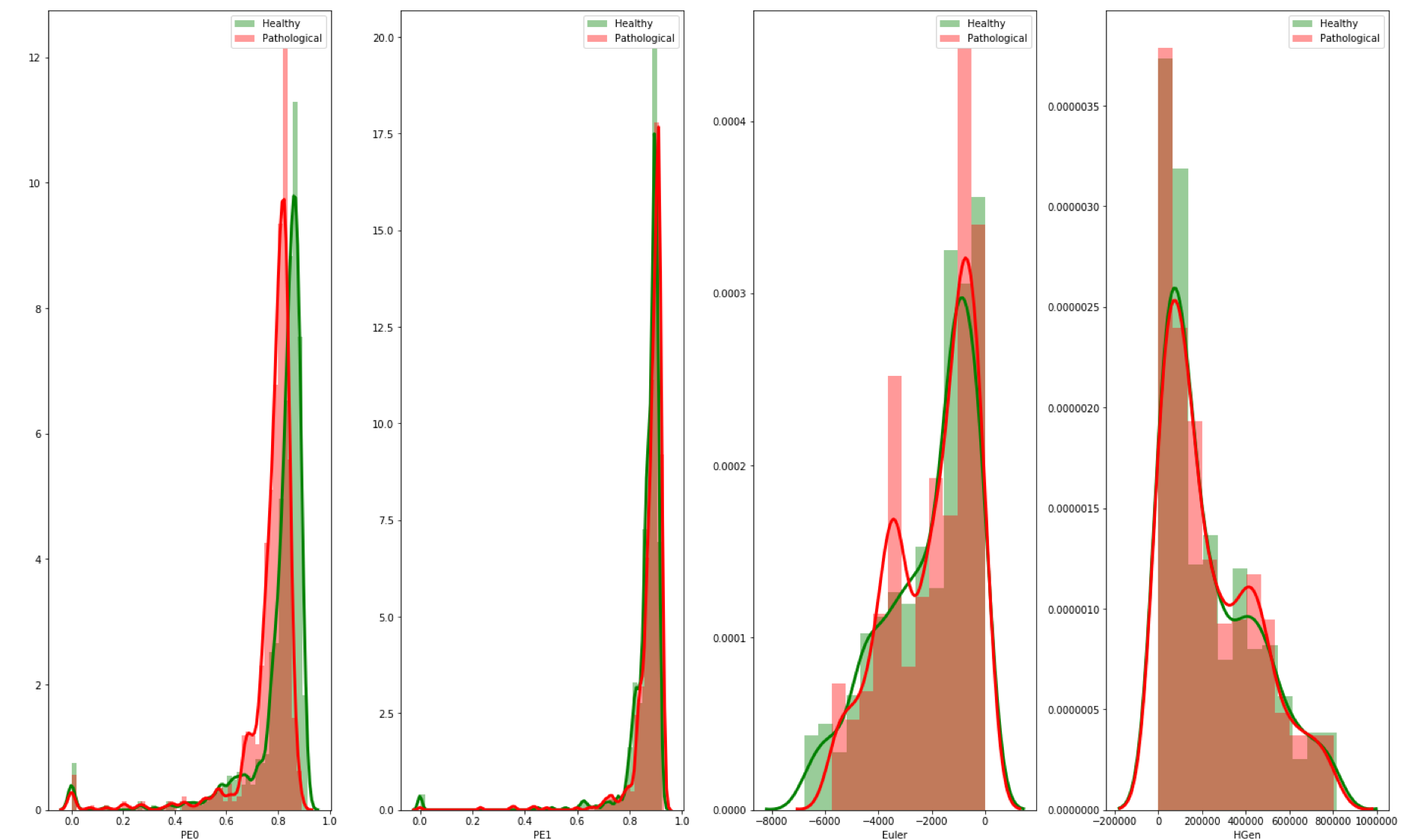}
\includegraphics[width=14cm]{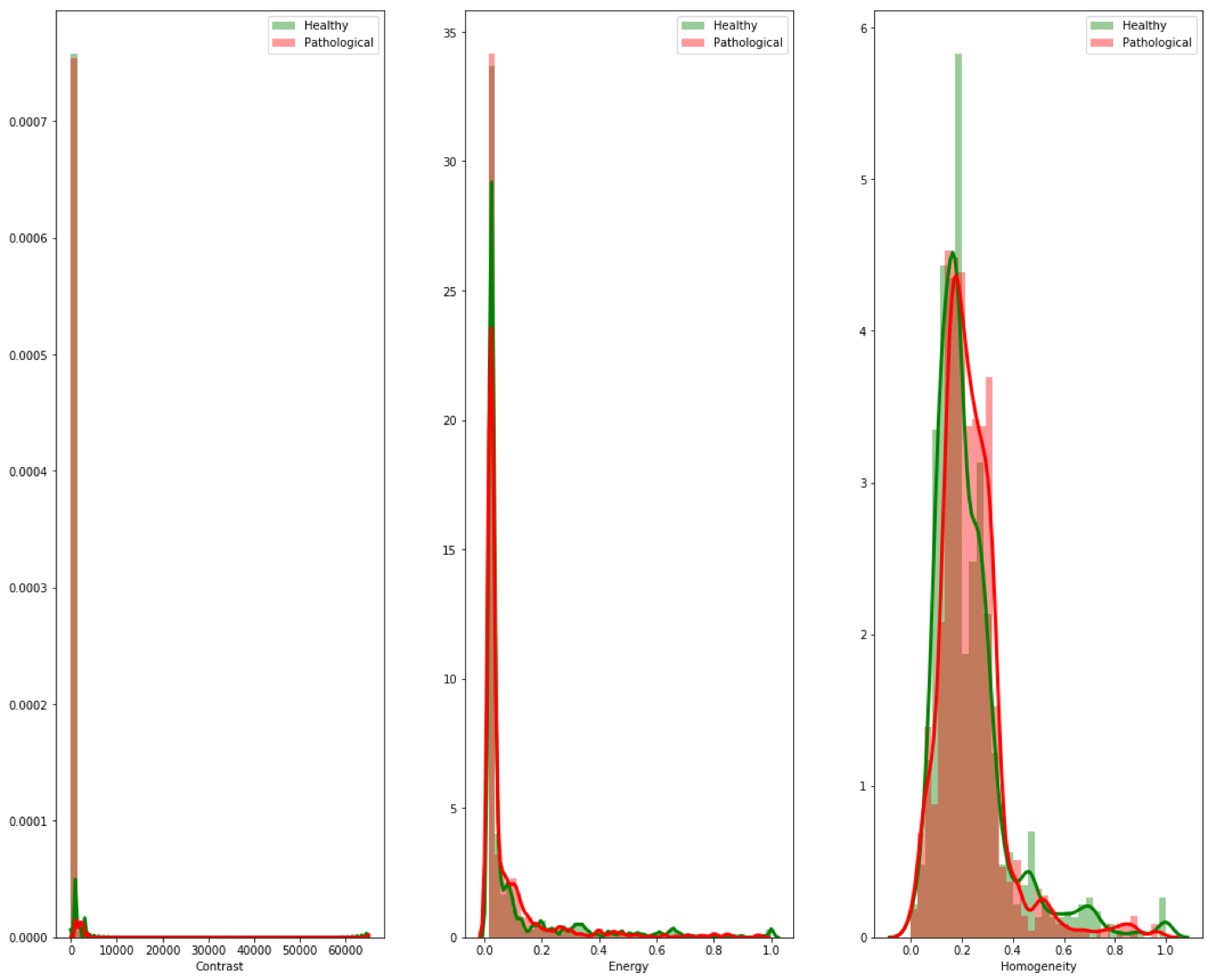}
\caption{Density distribution of the seven selected features: .}
\label{img:selection}
\end{figure}

Feature extraction comprises the process of extracting new features from the FLAIR. For each patch eight features (four topological plus four texture features) were computed. Features were standardized by removing the mean and scaling to unit variance\footnote{\url{https://scikit-learn.org/stable/modules/generated/sklearn.preprocessing.StandardScaler.html}}. Spearman's rank-order correlation combined with backward elimination was used to calculate the correlations between features and outcome variables.  Features with highest correlation and p-value$\leq0.05$ become more expressive of GBM characteristics. Fig.\ref{img:correlation} depicts the correlation among the features. Topological features are quite correlated among them. We observe that ''generator entropy" (HGen) shows a positive correlation with ''homogeneity".  This is in line with the idea that HGen is affected by how much the image is homogeneous both in terms of structure and of grey distribution. Eventually, the correlation analysis has selected seven out eight features. For the sake of completeness, all the topological features were selected while, the textural features \textit{correlation} was discarded.  Fig.\ref{img:selection} depicts the density distribution of the features that were selected. The topological features ''persistent entropy" for H0 and for H1 groups and the textural feature ''homogeneity"  show a quite good separation between ill (red) and healthy (green) group. The other features show different peaks but their distribution are overlapped.  

\subsubsection{Machine Learning algorithm selection}
The selection of ML algorithm was performed by using Automatic Machine Learning tool, i.e. Auto-Sklearn~\cite{NIPS2015_5872} and TPOT~\cite{olson2019tpot}. Auto-sklearn uses Bayesian optimization approaches for the automatic selection of machine learning classifier based on an ensemble of scikit-learn algorithms. Auto-sklearn compares and ranks different solutions and returns the solution tha maximizes the accuracy. TPOT is a framework that uses genetic algorithm for comparing different machine learning solutions and it finds automatically the best pipeline that maximizes the accuracy. Both Auto-Sklearn and TPOT produce a pipeline containing both preprocessing steps - if needed - and the selected Machine Learning architecture. 

\subsubsection{Deep Learning approach on numerical features}
For the sake of completeness, we have also trained deep learning classifiers on the selected features. In particular, we have trained multiple models with an architecture based on sequential dense layers with rectified linear unit (relu) activation function, followed by a dropout layer for preventing over-fitting and eventually by another dense layer as output layer. The difference among the models is the number of dense layers before the dropout and number of neurons per layers. 

\subsubsection{GBM classification -  Deep learning approach on patches}
\begin{figure}[!ht]
\centering
\includegraphics[width=14cm]{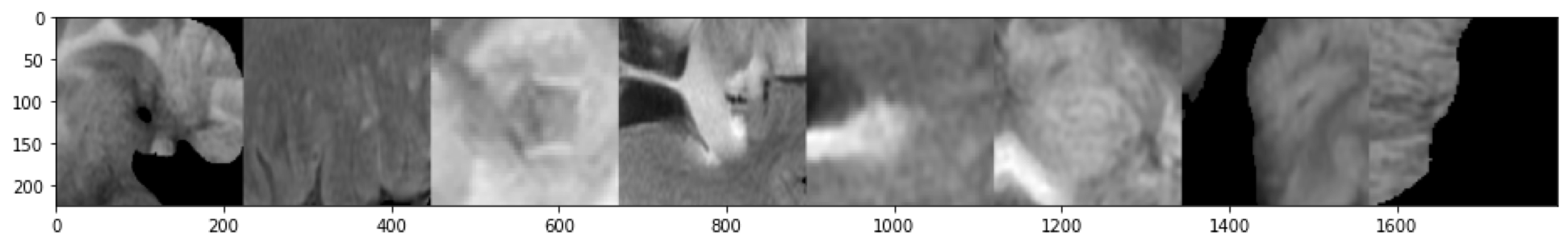}
\caption{Example of patches extracted from FLAIR.}
\label{img:patches}
\end{figure}

Deep-learning based solutions are opening new doors in the field of GBM detection and segmentation, we have compared the performance of the machine learning solution with two different Deep Neural networks on the patches extracted in the methodology defined in Sec.:\ref{sec:methodology}. Fig.:\ref{img:patches} contains 8 out of 1612 patches used for training and testing the algorithms. The first network contains 4 alternating convolutional and max-pooling layers, followed by a dropout after every other convolutional-pooling pair. After the last pooling layer, the network contains a fully-connected layer with 256 neurons, another dropout layer, then finally a softmax classification layer for two classes (i.e., pathological and healthy). The loss function is the categorical cross-entropy loss, and the learning algorithm is the AdaDelta. The network contains about 1 million parameters.  In the second setup, we have executed the transfer learning of a pre-trained VGG16 model. Transfer learning means the adaption of a pre-trained and optimized deep learning architecture for dealing with our dataset. The model consists of several convolutional layers, followed by some fully-connected / dense layers and then a softmax output layer for the classification. The dense layers are responsible for combining features from the convolutional layers. In this experiment, another dense-layer and a dropout-layer were added to avoid over-fitting.  VGG16 contains 13 convolutional layers and two fully connected layers at the end, and it counts more than 138 million parameters.  VGG16 was made to solve ImageNet, and achieves a $8.8\%$ top-5 error rate, which means that $91.2\%$ of test samples were classified correctly within the top 5 predictions for each image. The network was adapted to the two classes classification by  removing the original final classification layer, which corresponds to ImageNet, that was replaced by a new softmax layer which contains 2 neurons. Keras implementation of VGG16 facilitates to handle a such huge network and it optimizes memory issues\cite{zhao2018synthetic}. Both networks were trained  for 100 epochs with a batch size of 128.  For both networks  Keras API for TensorFlow and Google Colab environment\footnote{\url{https://colab.research.google.com/github/kylemath/ml4a-guides/blob/master/notebooks/transfer-learning.ipynb}} were used for the training.

\section{Results}
\label{sec:Results}

\subsection{Analysis 1: Topological Data Analysis of a simplified 2D tumour Growth Mathematical Model}
\begin{figure}[!ht]
\centering
\includegraphics[width=14cm]{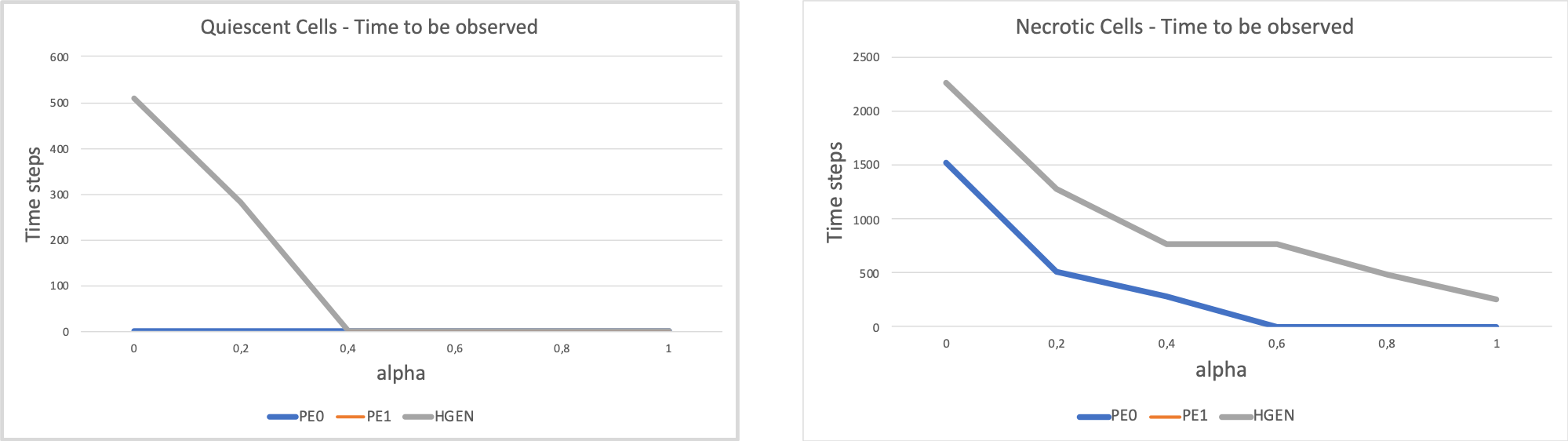}
\caption{Analysis of temporal properties of for quiescent and necrotic cells for different $\alpha=[0.0,0.5,1.0]$. Left: quiescent, right: necrotic. }
\label{img:temporal}
\end{figure}

Fig.:\ref{img:distribution} (top) depicts cells spatial displacement for $\alpha=[0.0,0.5,1.0]$ and the corresponding topological statistics (bottom) computed throughout system integration. At first look, with low $\alpha$ the quiescent and necrotic cells (in blue and black respectively) are quite confined in the central region, while for high $\alpha$ they are more scattered. We have further investigated this aspect by plotting for different $\alpha$ the amount of time steps that were needed to find topological features greater than zero. Since proliferative cells are always present and thus the corresponding topological features are immediately greater than zero, we have focused our analysis only on quiescent and necrotic cells. The outcome of the analysis is represented in Fig.:\ref{img:temporal}. At high concentration of nutrients $\alpha \geq 0.4$ the phenomenon is more virulent and therefore it takes less time for proliferating both quiescent and necrotic cells.

\subsection{Analysis 2: Topological Data Analysis of tumour Progression}
Topological features detect the change of brain shape between the ''pre" and ''post" CRT treatment. Fig.:\ref{img:distribution} describes the density estimation for the topological features (Euler Characteristics, Persistent Entropy for Ho and H1 and Generator Entropy). It is well evident that the two distributions of persistent entropy for the connected components (H0) are quite different: the distribution for the ''pre" set (red) is  bimodal with a local maximum at 0.0 and a global maximum $\approx$ at 0.6. The distribution for the ''post" set (blue) contains two local maxima and a global maximum $\approx$ at 0.7. Welch's t-test was computed for testing the null hypothesis that the means of the distribution are equal. The outcome of the analysis is capture in Tab.:\ref{tab:tumourProgression}. Only the distributions for the peristent entropy at H0 are statistically different. This result would suggest to use persistent entropy as a new measure for monitoring the follow-up of chemo-radiation therapy (CRT) treatment for GBM.

\begin{table}[!ht]
\begin{tabular}{|l|l|l|l|l|}
\hline
\multicolumn{5}{|c|}{Topological Statistics}          \\ \hline
                      & Pre           & Post & t-test & p-value          \\ \hline
Euler Characteristics & -773.94  +/- 62.98 & -830.31 +/- 74.55 & 1.84 & 0.06 \\ \hline
Persistent Entropy H0 & 0.44 +/- 0.23 & 0.51 +/- 0.24 & -6.79 & \textbf{0.01} \\ \hline
Persistent Entropy H1 & 0.70 +/- 0.15 & 0.70 +/- 0.16 & -0.29 & 0.77 \\ \hline
Generator Entropy H1  & 0.30 +/- 0.26 & 0.28 +/- 0.27 & 1.68 & 0.09 \\ \hline
\end{tabular}
\caption{Statistical analysis of topological features for tumour progression analysis.}
\label{tab:tumourProgression}
\end{table}

\begin{figure}[!ht]
\centering
\includegraphics[width=14cm]{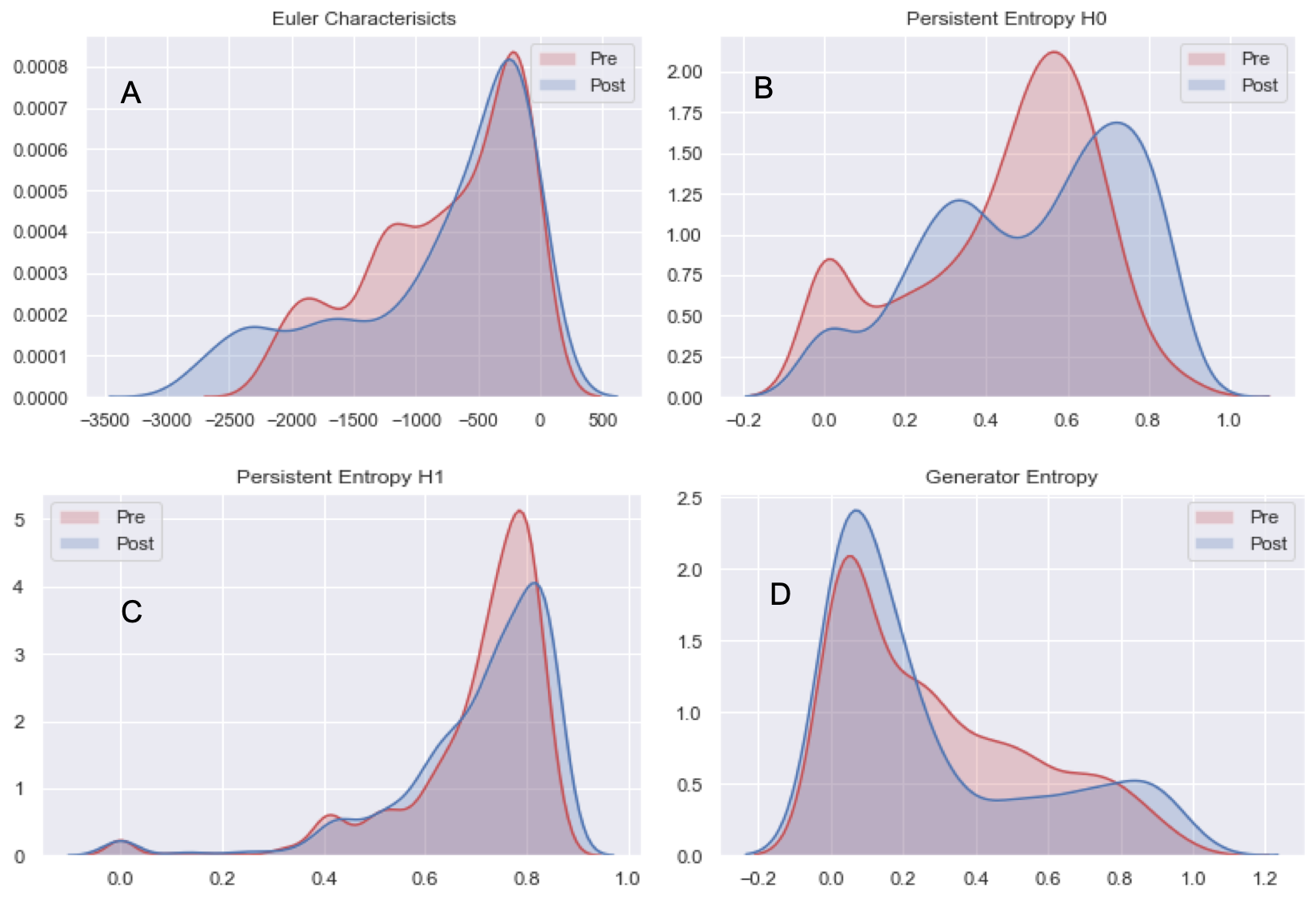}
\caption{Density estimation for the topological features: comparison between the features computed on the ''pre" and on the ''post" set of 2D FLAIR.}
\label{img:distribution}
\end{figure}

\subsection{Analysis 3: GBM classification on FLAIR}
\begin{figure}[!ht]
\centering
\includegraphics[width=8cm]{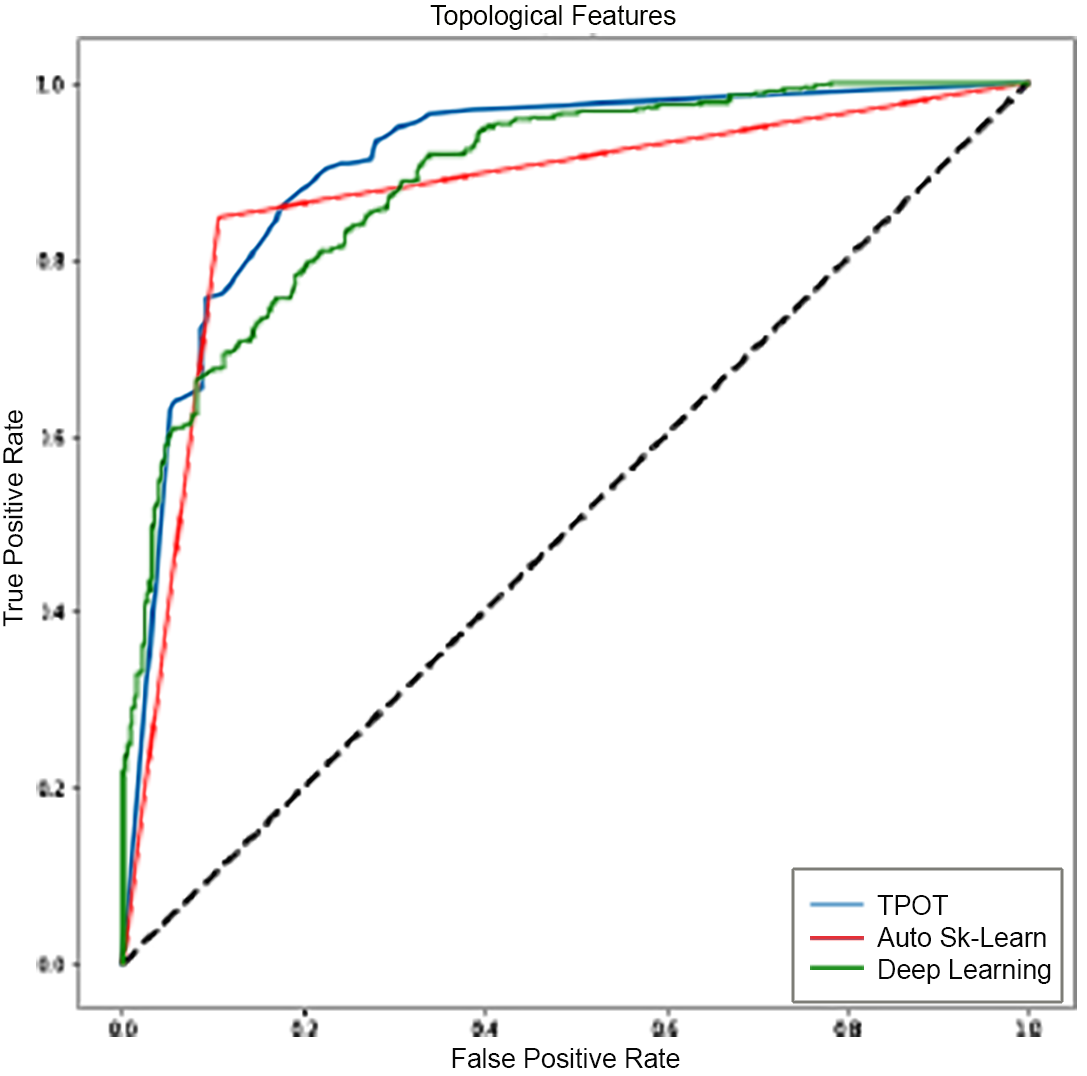}
\caption{Comparison of ROC curves for the three machine learning approaches trained with topological features.}
\label{img:ROC_Topological}
\end{figure}

\begin{figure}[!ht]
\centering
\includegraphics[width=8cm]{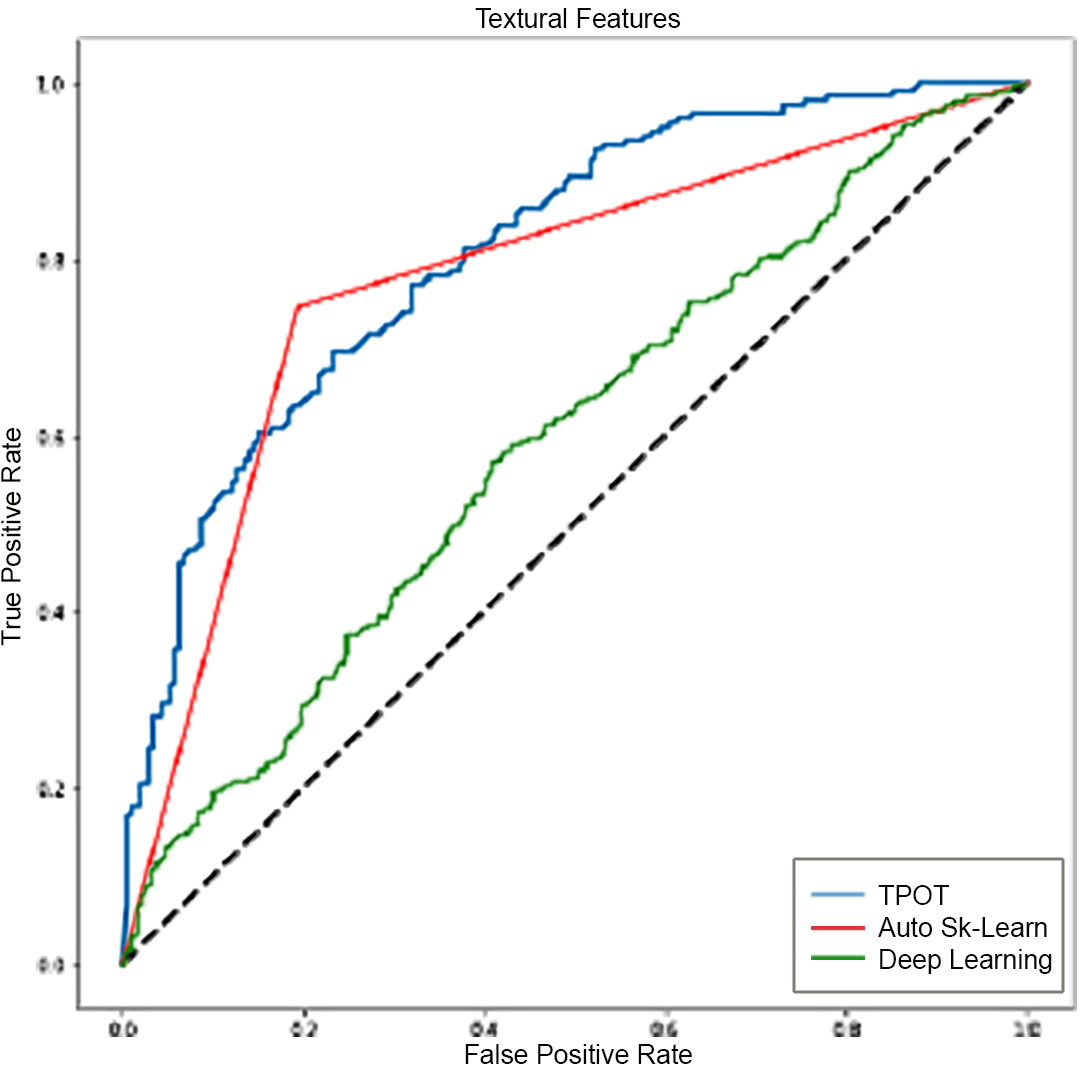}
\caption{Comparison of ROC curves for the three machine learning approaches trained with textural features.}
\label{img:ROC_Textural}
\end{figure}

\begin{figure}[!ht]
\centering
\includegraphics[width=8cm]{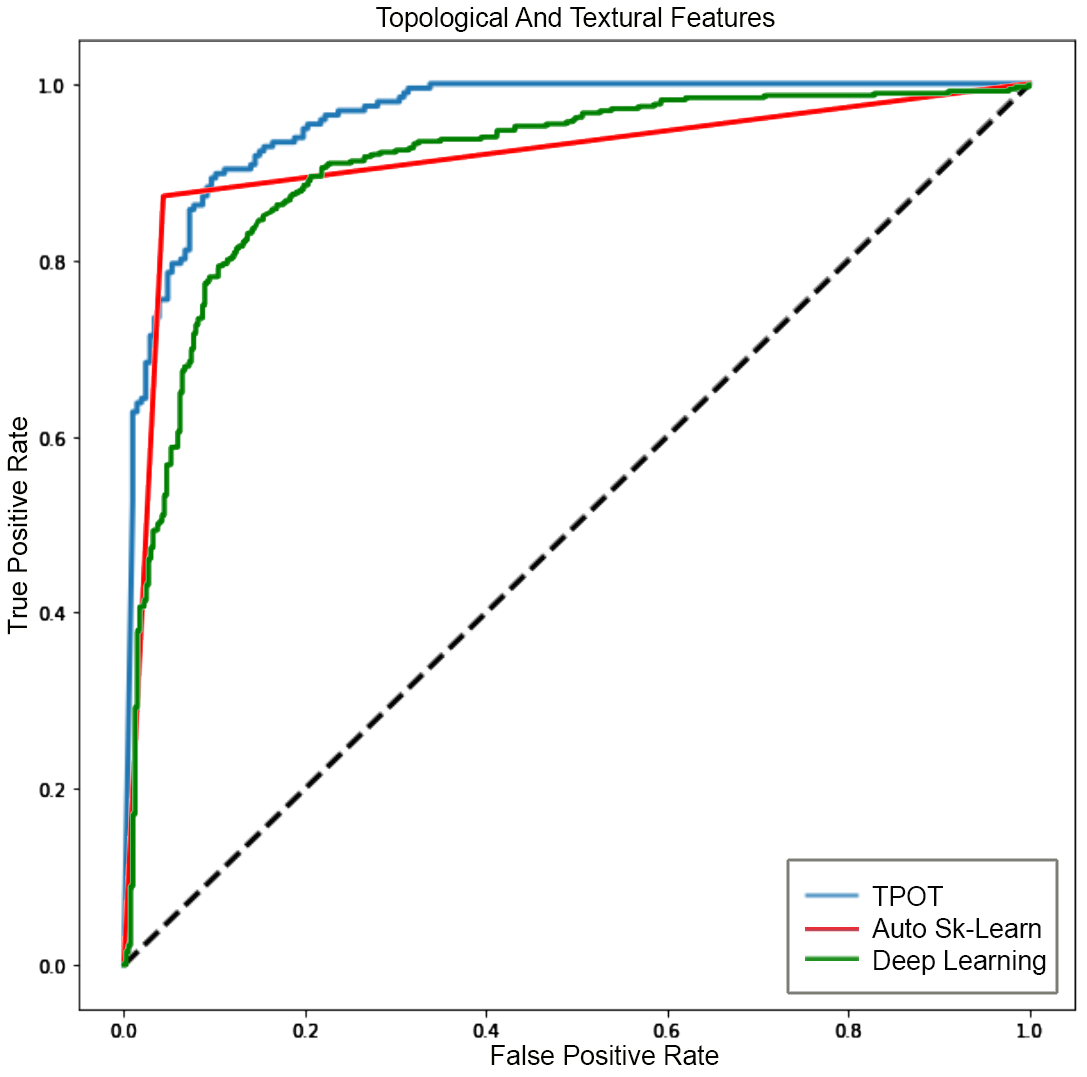}
\caption{Comparison of ROC curves for the three machine learning approaches trained with topological and textural features.}
\label{img:ROC_TopoText}
\end{figure}

Tables \ref{tab:Topo}, \ref{tab:Text} and \ref{tab:TopoTex} show the results obtained in the training and test sets respectively, grouped by the algorithms testes in this study. The corresponding ROC curves are depicted in Fig.:\ref{img:ROC_Topological}, \ref{img:ROC_Textural} and \ref{img:ROC_TopoText}.  In general, TPOT produces the best algorithms in combination with the proposed preprocessing steps. The worse performances are obtained by using only the textural features. The combination of topological and textural features rise the best performances. In both cases, namely only topological and topological plus textural, the accuracy are quite good.  We argue that TPOT shows better performance because it optimizes over more parameters increase the separation of the features with respect the two classes~\cite{ahlgren2018auto}. 

\begin{table}[!ht]
\begin{tabular}{|l|l|l|l|l|l|l|}\hline
\multicolumn{7}{|c|}{Topological Features}\\ \hline
         & \multicolumn{2}{|c|}{TPOT} & \multicolumn{2}{|c|}{Auto Sci-Kit learn} & \multicolumn{2}{|c|}{Deep Learning} \\ \hline
        Metric                  & Train  &      Test  & Train  &      Test  & Train  &      Test  \\ \hline
        Accuracy                & 0.89 &  0.84  & 0.97  &  0.87  & 0.82  &  0.79   \\ \hline
        Precision               & 0.88 &  0.83  & 0.97  &  0.88  & 0.90  &  0.91  \\ \hline
        Recall                  & 0.88 &  0.84  & 0.96  &  0.84  & 0.72  &  0.67  \\ \hline
        Misclassification rate  & 0.10 &  0.16  & 0.03  &  0.13  & 0.17  &  0.20   \\ \hline
        F1                      & 0.89 &  0.83  & 0.97  &  0.88  & 0.80  &  0.77  \\ \hline
        AUC                     & 0.89 &  0.89  & 0.97  &  0.87  & 0.82  &  0.79  \\ \hline
        \end{tabular}
\caption{Machine Learning Accuracy - Topological Features}
\label{tab:Topo}
\end{table}

\begin{table}[!ht]
\begin{tabular}{|l|l|l|l|l|l|l|}\hline
\multicolumn{7}{|c|}{Textural Features}\\\hline
        & \multicolumn{2}{|c|}{TPOT} & \multicolumn{2}{|c|}{Auto Sci-Kit learn} & \multicolumn{2}{|c|}{Deep Learning} \\ \hline
        Metric                  & Train  &      Test  & Train  &      Test  & Train  &      Test  \\ \hline
        Accuracy                & 1.00 &  0.71  & 0.93   &  0.77  & 0.60  &  0.58  \\ \hline
        Precision               & 1.00 &  0.70  & 0.94   &  0.78  & 0.61  &  0.60  \\ \hline
        Recall                  & 1.00 &  0.71  & 0.92   &  0.74  & 0.51  &  0.51   \\ \hline
        Misclassification rate  & 0.00 &  0.28  & 0.07   &  0.23  & 0.40  &  0.42  \\ \hline
        F1                      & 1.00 &  0.71  & 0.93   &  0.76  & 0.56  &  0.55  \\ \hline
        AUC                     & 1.00 &  0.81  & 0.93   &  0.77  & 0.60  &  0.57  \\ \hline\end{tabular}
\caption{Machine Learning Accuracy - Textural Features}
\label{tab:Text}
\end{table}

\begin{table}[!ht]
    \begin{tabular}{|l|l|l|l|l|l|l|}\hline
        \multicolumn{7}{|c|}{Topological and Textural Features}\\\hline
        & \multicolumn{2}{|c|}{TPOT} & \multicolumn{2}{|c|}{Auto Sci-Kit learn} & \multicolumn{2}{|c|}{Deep Learning} \\ \hline
         Metric                  & Train  &      Test  & Train  &      Test  & Train  &      Test  \\ \hline
        Accuracy                & 0.96  &  0.89  &  0.98  &  0.92   & 0.90  &  0.89  \\ \hline
        Precision               & 0.96  &  0.89  &  0.99  &  0.95   & 0.87  &  0.85   \\ \hline
        Recall                  & 0.95  &  0.89  &  0.97  &  0.87   & 0.93  &  0.84  \\ \hline
        Misclassification rate  & 0.04  &  0.10  &  0.02  &  0.08   & 0.09  &  0.15  \\ \hline
        F1                      & 0.96  &  0.89  &  0.98  &  0.91   & 0.90  &  0.84  \\ \hline
        AUC                     & 0.99  &  0.96  &  0.98  &  0.91   & 0.90  &  0.84  \\ \hline
    \end{tabular}
\caption{Machine Learning Accuracy - Topological and Textural Features}
\label{tab:TopoTex}
\end{table}

\subsubsection{Machine Learning classification interpretation}
\begin{figure}[!ht]
\centering
\includegraphics[width=8cm]{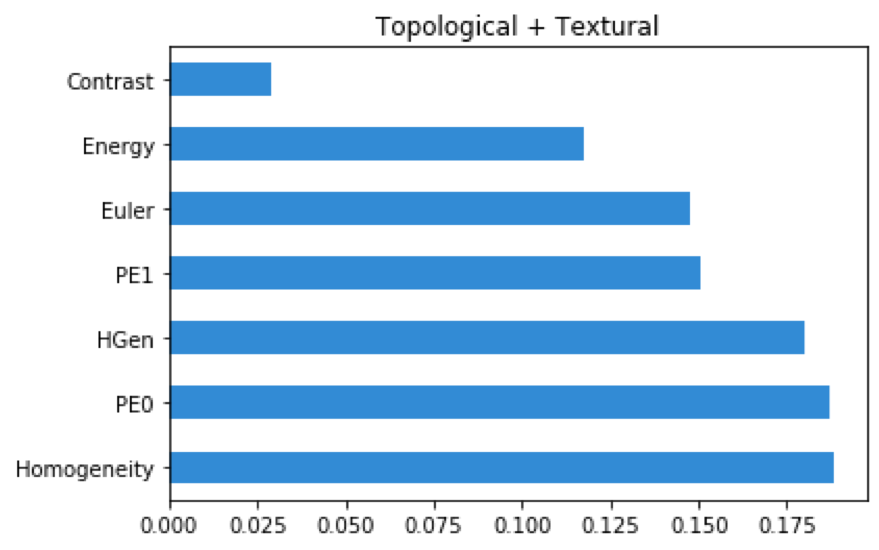}
\caption{Global features relevance analysis w.r.t TPOT prediction.}
\label{img:skater}
\end{figure}

Skater and Lime algorithms were used for computing features relevance with respect to the TPOT machine learning algorithms trained on topological and textural features~\cite{wei2015variable}. The output of Skater is depicted in Fig.:\ref{img:skater}. 
For TPOT classifier, the textural features \textit{homogeneity} has the same relevance of the topological entropy \textit{persistent entropy at H0 - PE0} that is followed by  \textit{generator entropy - HGEN}. This confirms that the combination of topological and textural features allowed to reach high accuracy.  The less relevant features is \textit{contrast}, however since its importance is greater than 0 it cannot be discarded.

\begin{figure}[!ht]
\centering
\includegraphics[width=8cm]{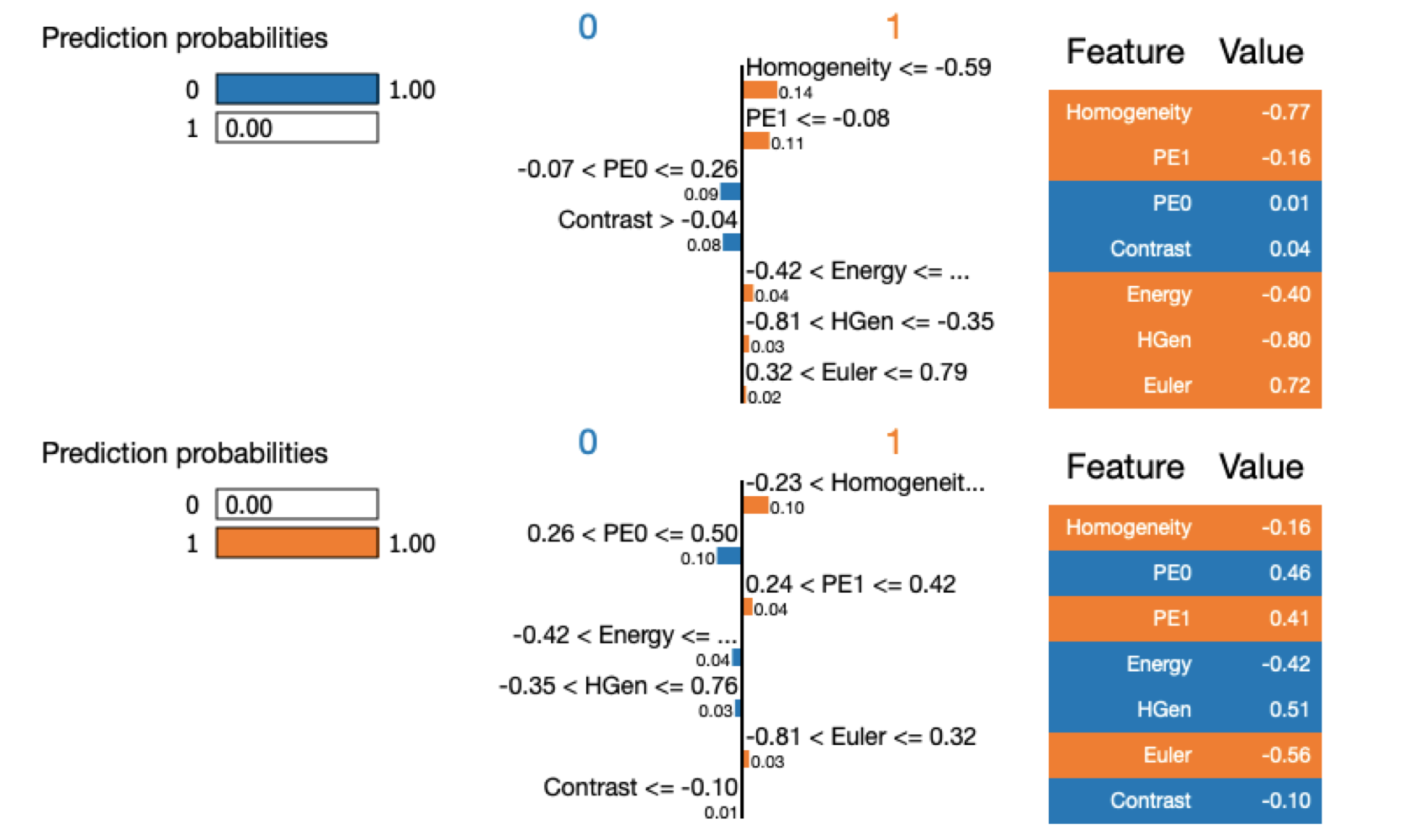}
\caption{Example of local features relevance analysis for two different samples in the test set w.r.t TPOT prediction.}
\label{img:lime}
\end{figure}

The output of Lime is represented in Fig.:\ref{img:lime}. Lime algorithm has allowed to understand what are the numerical characteristic for a patch to be classified healthy or ill. Negative (blue) indicate healthy tissues, while positive (orange) indicate ill tissue. The way to interpret the weights by applying them to the prediction probabilities is straightforward. The bars indicate the weight for each feature and the corresponding value for the slice under test. The subtraction of the weights from the prediction probabilities (1 in both cases as indicated in the left part of the picture) will alter the probability of a sample to be classified ill or healthy. This highlights that even if some features might appear globally less relevance, they are locally fundamental for avoiding ''flipping coin" predictions. 

\subsection{GBM classification with Deep Learning}
\begin{figure}[!ht]
\centering
\includegraphics[width=8cm]{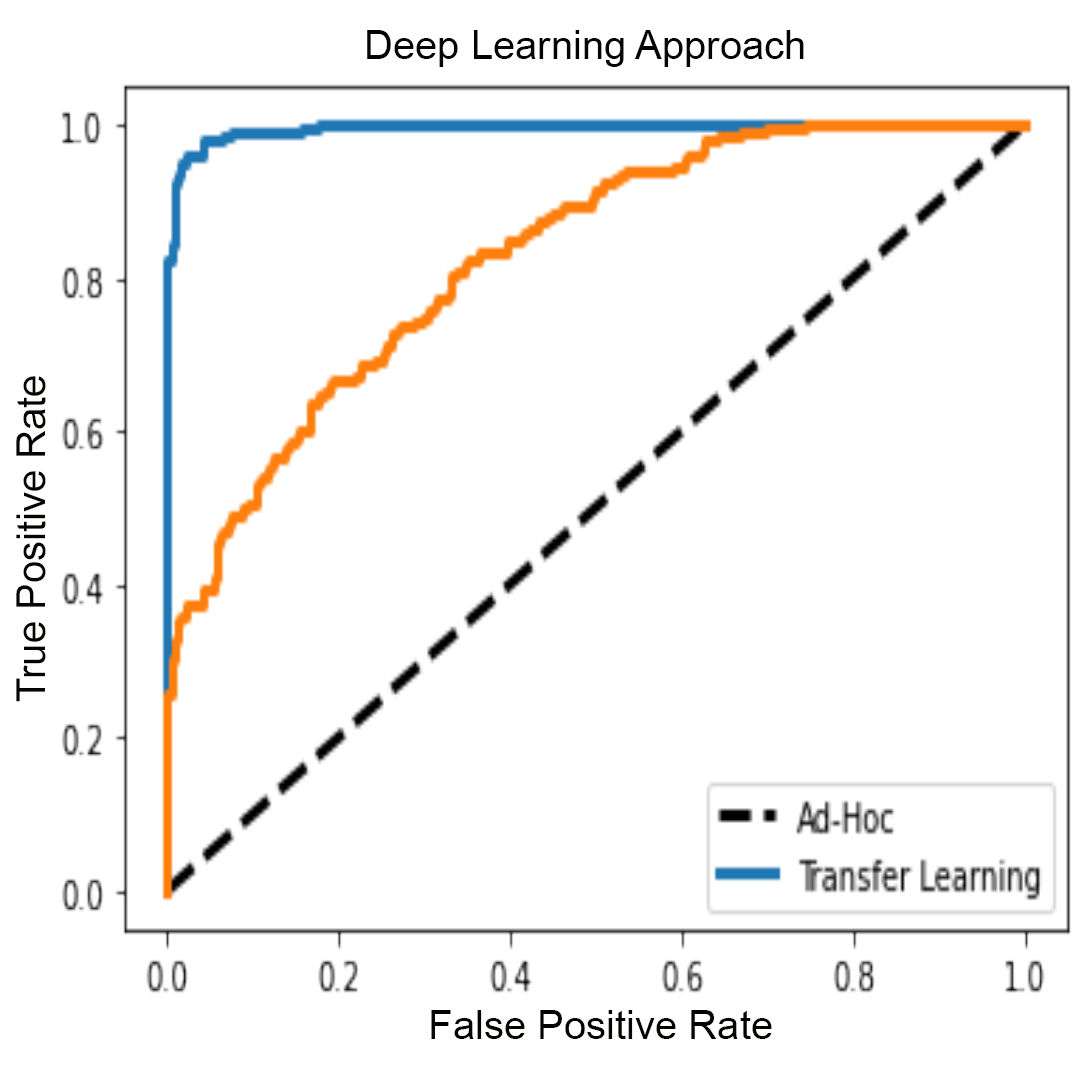}
\caption{Comparison of ROC curves for the two deep-learning based approaches.}
\label{img:ROC_Deep_Learning}
\end{figure}

\begin{figure}[!ht]
\centering
\includegraphics[width=8cm]{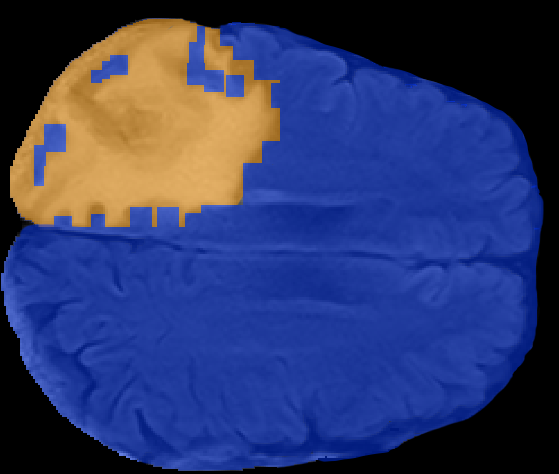}
\caption{Examples of patches classification with fine-tuned VGG16: blue patches correspond to healthy tissue, orange patches correspond to ill tissue. The resolution of the segmentation is a consequence of rectangular shapes of the patches.}
\label{img:segmentation}
\end{figure}

Table \ref{tab:DeepLearning} shows the results obtained in the training and test sets respectively, grouped by the deep learning algorithms tested in this study. The corresponding ROC curves are depicted in Fig.:\ref{img:ROC_Deep_Learning}.  The fine-tuning of VGG16 with the transfer-learning approach has allowed to reach the best performances. By comparing the metrics (e.g., accuracy, precision-recall, misclassification rate and F1) it is evident that VGG16 outperforms TPOT. Only the AUC of TPOT ($96\%$) is quite closed to the AUC reached by VGG16 ($97\%$). Finally, examples of patches classification achieved by applying the fine-tuned VGG16 on an entire 2D FLAIR is showed in Fig.:\ref{img:segmentation}.

\begin{table}[!ht]
\begin{tabular}{|l|l|l|l|l|}
\hline
\multicolumn{5}{|c|}{Deep Learning on Patches}                                                      \\ \hline
                       & \multicolumn{2}{|c|}{VGG16 Transfer Learning} & \multicolumn{2}{|c|}{Ad-Hoc} \\ \hline
Metric                 & Train         & Test        & Train                 & Test                 \\ \hline
        Accuracy                & 0.99   &  0.97  & 0.74  &  0.75  \\ \hline
        Precision               & 1.00 &  0.99   & 0.67  &  0.66  \\ \hline
        Recall                  & 0.98   &  0.95   & 0.97   &  0.95  \\ \hline
        Misclassification rate  & 0.01  &  0.03  & 0.26  &  0.25  \\ \hline
        F1                      & 0.99     &  0.97   & 0.79  &  0.78  \\ \hline
        AUC                     & 0.99   &  \textbf{0.97}   & 0.74 &  0.77  \\ \hline
\end{tabular}
\caption{Deep Learning Accuracy using patches}
\label{tab:DeepLearning}
\end{table}

\section{Discussion}
\label{sec:conclusion}
Topological data analysis is a reliable approach to shed light on GBM characteristics, as it has demonstrated by the three numerical experiments described in this paper. Fast and reliable assessment of GBM presence is critical for an accurate surgical and radiotherapy planning as well as to evaluate and quantify treatment-related effects and progression In the recent period, surgical approaches are strongly changed, with a progressive shift from classical neurosurgical techniques towards newer approaches with surgical navigation systems fluorescence-guided and MR-guided surgery ~\cite{weller2014european, stummer2006fluorescence, kuhnt2011correlation}. Also, chemotherapy employed techniques with wafer drugs positioning in situ to minimize systemic collateral effects~\cite{hart2008chemotherapeutic}. Consequently, radiological techniques able to localize and evidence every minimum GBM localization are essential to enforce new therapeutic possibilities able to improve the extent of tumour resection, prolong survival and increase the quality of life.  
Topological descriptor such as Euler Characteristics, Persistent Entropy and Persistent Generator were used to identify the physiological conditions that facilitate tumour growth. We notice that \textit{Generator entropy} was introduced in this paper as a novel topological statistic that summarizes the number of homological generators of nD homological loops. In particular, topological statistics were able to identify the change of  proliferation rate of quiescent and necrotic cells as a consequence of nutrient concentration: the higher the concentration of chemical nutrients the more virulent the process. We would to remark once again that in this paper we had analysed a simplified 2D tumour growth model, in order to develop a more reliable personalized tumour growth analysis we envision the need to combine topological dynamical system analysis with imaging analysis~\cite{roose2007mathematical,bender2016pt}. 
In the other two experiments we have demonstrated by using a publicly available dataset that topological features can be used as new radiomics features for GBM characterization. The analysis of FLAIR sequences recorded within 90 days following chemo-radiation therapy (CRT) completion and at progression by means of topological descriptors has confirmed by statistical test that persistent entropy computed at the homological group H0 of the connected component is a viable alternative for monitoring treatment effects.
Based on these results we had proposed a new method for GBM detection and segmentation. The supervised method for automatic classification of 2D patches extracted from FLAIR sequences is a viable alternative for GBM identification and segmentation.  The proposed method comprises the following main stages: image preprocessing, sliding patches extraction, topological and textural features computation and selection, supervised classification via automatic machine learning, interpretation of classifier prediction. Preprocessing steps, i.e. skull stripping,  were obtained by employing the latest deep learning technology. Several supervised classification algorithms were evaluated to ensure the highest classification accuracy.  To this extent, two different automatic machine learning selection frameworks were evaluated, which are TPOT and Auto-SkLearn. The algorithms were trained using  topological features or textural features or the combination of the two sets. The selected machine learning approaches were used as baseline in the comparison with deep learning based approaches. For the sake of completeness, deep learning approaches were trained on the same features set or directly on the patches (sub-images) extracted from the 2D FLAIR. The quality of classification was assessed by several metrics and ROC curves. The highest accuracy using the features was reached by TPOT trained on the two sets (e.g., topological and textural) with an accuracy of $89\%$ and $AUC=96\%$. Finally, the trained systems were investigated with tools from information theory. Skater and Lime algorithms allowed computing features relevance and for understanding what are the numerical characteristics of a patch to be classified ill or healthy. The interpretation has confirmed the importance of topological features. The transfer learning of a pre-trained VGG16 network allowed for patches classification had allowed to reach an accuracy of $97\%$ and $AUC=97\%$. However, we would like to remark that in general Deep Learning based methods require ad-hoc hardware for their training, in fact we have used Google Colab platform that allows to access to both GPU and TPU units.
The adoption of deep learning solution would be straightforward, but we remark that machine learning classification verdicts can be easily interpreted by enrolling tools like Skater and Lime while, the interpretation of the feature space extracted by a deep learning solution might be unfeasible, since the convolutional features might not correspond to any physical information.
In the future, we envision several efforts: we intend to investigate mathematical properties (minimum set of generators, stability theorem, etc\dots) of the new topological entropy~\cite{obayashi2018volume}. We intend also to extend and make more precise our study by extending the cohort, for example by including FLAIR from  other datasets (e.g., BRATS), and by challenging U-Net algorithm and to equip it with tools for outcome interpretation. Our end will be to develop a novel Computer-Aided Detection tool compliant with the EU-GDPR 22nd article (''Automated individual decision-making, including profiling") for the ''right to be informed"\footnote{\url{http://www.privacy-regulation.eu/it/22.htm}}.\\ 

\authorcontributions{M.R. has conceptualized the methodology and executed the experiment. G.V. and L.F. have contributed to paper writing.}

\funding{``This research received no external funding''}

\acknowledgments{The author would like to thank Prof. Rocio Diaz-Gonzalez and Prof. Matthias Zeppelzauer for fruitful discussions on topological statistics for imaging;  Dr. Giorgio De Nunzio, Dr. Marina Donativi and MD Antonella Castellano for long-standing conversations on GBM segmentation. Andrea De Antoni for the discussions on how to improve the presentation of the main findings.}

\conflictsofinterest{The authors declare no conflict of interest.} 

\abbreviations{The following abbreviations are used in this manuscript:\\

\noindent 
\begin{tabular}{@{}ll}
GBM & Glioblastomas multiforme\\
GLCM & grey Level Co-occurrence Matrix\\
TDA & Topological Data Analysis\\
MRI & Magnetic Resonance Image\\
FLAIR & Fluid-attenuated inversion recovery\\
MD & Medical Doctor\\
\end{tabular}}

\reftitle{References}


\externalbibliography{yes}
\bibliography{paperRV}  





\end{document}